\documentclass{PoS}
\def\krig#1{\vbox{\ialign{\hfil##\hfil\crcr
           $\raise0.3pt\hbox{$\scriptstyle \circ$}$\crcr\noalign
           {\kern-0.02pt\nointerlineskip}
$\displaystyle{#1}$\crcr}}}

\newcommand{\fo}{\krig{f}}
\newcommand{\go}{\krig{g}}
\newcommand{\xo}{\krig{\langle x \rangle}}

\title{Generalized parton distributions from domain wall valence quarks and staggered sea quarks}

\ShortTitle{Generalized parton distributions from domain wall valence quarks and staggered sea quarks}

\author{LHPC Collaboration:\ \speaker{D.B.~Renner},$^a$\hspace{-1pt}\thanks{Current address:\ DESY Zeuthen,
Platanenallee 6, D-15738 Zeuthen, Germany}\hspace{2pt} J.~Bratt,$^b$ R.G.~Edwards,$^c$ M.~Engelhardt,$^d$
G.~Fleming,$^e$ Ph.~H\"agler,$^f$ B.~Musch,$^f$ J.W.~Negele,$^b$ K.~Orginos,$^{cg}$ A.V.~Pochinsky,$^b$
D.G.~Richards,$^c$ W.~Schroers$^h$\\
\llap{$^a$}University of Arizona, Department of Physics, 1118 E 4th St, Tucson AZ 85721, USA\\
\llap{$^b$}Center\hspace{-0.0725pc} for\hspace{-0.0725pc} Theoretical\hspace{-0.0725pc} Physics,\hspace{-0.0725pc} Massachusetts\hspace{-0.0725pc} Institute\hspace{-0.0725pc} of\hspace{-0.0725pc} Technology,\hspace{-0.0725pc} Cambridge,\hspace{-0.0725pc} MA\hspace{-0.0725pc} 02139,\hspace{-0.0725pc} USA\\
\llap{$^c$}Thomas Jefferson National Accelerator Facility, Newport News, VA 23606, USA\\
\llap{$^d$}Physics Department, New Mexico State University, Las Cruces, NM 88003-8001\\
\llap{$^e$}Sloane Physics Laboratory, Yale University, New Haven, CT 06520, USA\\
\llap{$^f$}Institut f\"ur Theoretische Physik, TU M\"unchen, D-85747 Garching, Germany\\
\llap{$^g$}Department of Physics, College of William and Mary, Williamsburg VA 23187, USA\\
\llap{$^h$}John von Neumann-Institut f\"ur Computing NIC/DESY, D-15738 Zeuthen, Germany}

\abstract{Moments of the generalized parton distributions of the nucleon, calculated with
a mixed action of domain wall valence quarks and asqtad staggered sea quarks,
are presented for pion masses extending down to 359 MeV. Results for the
moments of the unpolarized, helicity, and transversity distributions are given
and compared to the available experimental measurements. Additionally, a
selection of the generalized form factors are shown and the implications for
the spin decomposition and transverse structure of the nucleon are discussed.
Particular emphasis is placed on understanding systematic errors in the lattice
calculation and exploring a variety of chiral extrapolations.}

\FullConference{The XXV International Symposium on Lattice Field Theory\\July 30 - August 4 2007\\Regensburg, Germany}

\begin{document}

\section{Introduction}

\begin{figure}
\begin{minipage}{17pc}\vspace{1.9pc}
\begin{center}
\begin{tabular}{|l|l|l|l|l|} \hline
$a m_{u/d}^{\mathrm{asqtad}}$ & $L/a$ & $L$ & $m_\pi^{\mathrm{DWF}}$ & \# \\ \hline
 & & $\mathrm{fm}$ & $\mathrm{MeV}$ & \\ \hline
$0.05$ & $20$ & $2.52$ & $761$ & $425$ \\ \hline
$0.04$ & $"$  & $"$    & $693$ & $350$ \\ \hline
$0.03$ & $"$  & $"$    & $594$ & $564$ \\ \hline
$0.02$ & $"$  & $"$    & $498$ & $486$ \\ \hline
$0.01$ & $"$  & $"$    & $354$ & $656$ \\ \hline
$0.01$ & $28$ & $3.53$ & $353$ & $270$ \\ \hline
\end{tabular}
\end{center}\vspace{-0.5pc}
\caption{\label{param}Lattice parameters used in this work.}
\end{minipage}
\begin{minipage}{17pc}
\begin{center}
\includegraphics[width=16pc,clip]{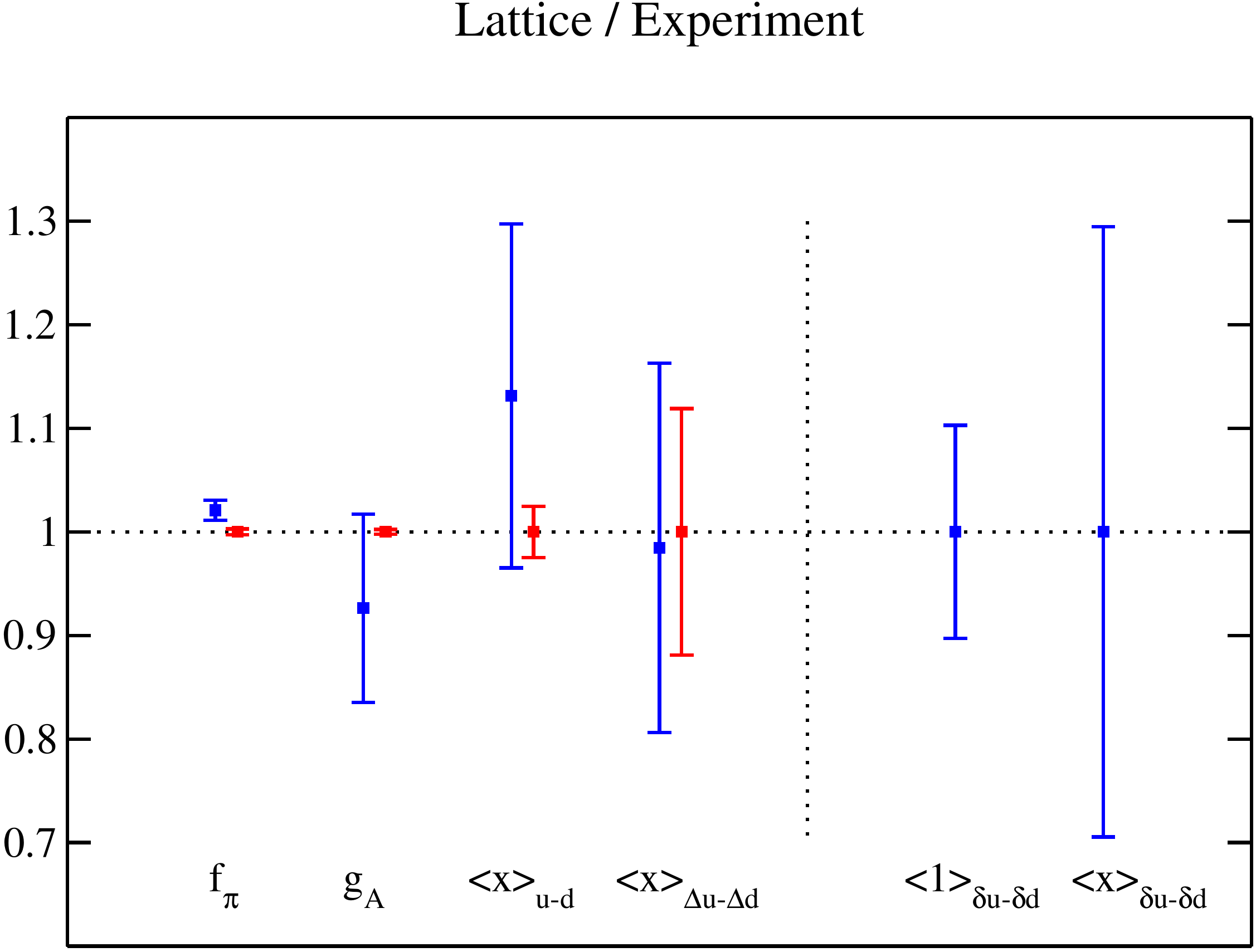}
\caption{\label{sum}Moments of parton distributions.}
\end{center}
\end{minipage}
\end{figure}

Lattice field theory provides a precise definition of nucleon
structure observables as well as a numerical means for evaluating them
non-perturbatively.  Here we focus on the nucleon generalized parton
distributions.  This set of observables contains the parton
distributions and form factors as well as quantities that determine
the transverse distribution of quarks within the nucleon and the
decomposition of the nucleon spin into quark and gluon degrees of
freedom.  To perform these calculations, we use a mixed action
consisting of domain wall valence quarks on improved staggered sea
quark configurations provided by the MILC
collaboration~\cite{Bernard:2001av}.  The details of this work have
been published in Refs.~\cite{Renner:2004ck}, \cite{Edwards:2005kw},
and \cite{Edwards:2006qx} and are briefly summarized in
Fig.~[\ref{param}].  In this proceeding, we examine the systematic
errors related to the determination of matrix elements from lattice
correlation functions as well as the extrapolation of lattice results
to the physical quark masses.

\section{Moments of Parton Distributions}

As theorists improve algorithms and as the computing resources
dedicated to lattice QCD calculations continue to grow, we can begin
to contemplate precise calculations of the low non-singlet moments of
the nucleon parton distributions.  This will require control of a
variety of sources of error.  The calculation of any observable
requires a study of the volume, lattice spacing, and quark mass
dependence.  Additional sources of error arise depending on the
observable and choice of action.  For the calculation of moments of
parton distributions, the additional errors are caused by matching
correlation functions calculated on the lattice to functional forms
derived or motivated by transfer matrix arguments as well as errors
due to the renormalization of the operators themselves.  Our previous
work examined the volume and quark mass dependence of the axial
charge~\cite{Edwards:2005ym} and the quark mass dependence of all the
low moments~\cite{Edwards:2006qx} and the generalized form
factors~\cite{Hagler:2007xi}.

\subsection{Correlation Functions}

\begin{figure}
\begin{minipage}{17.75pc}
\includegraphics[width=16.5pc]{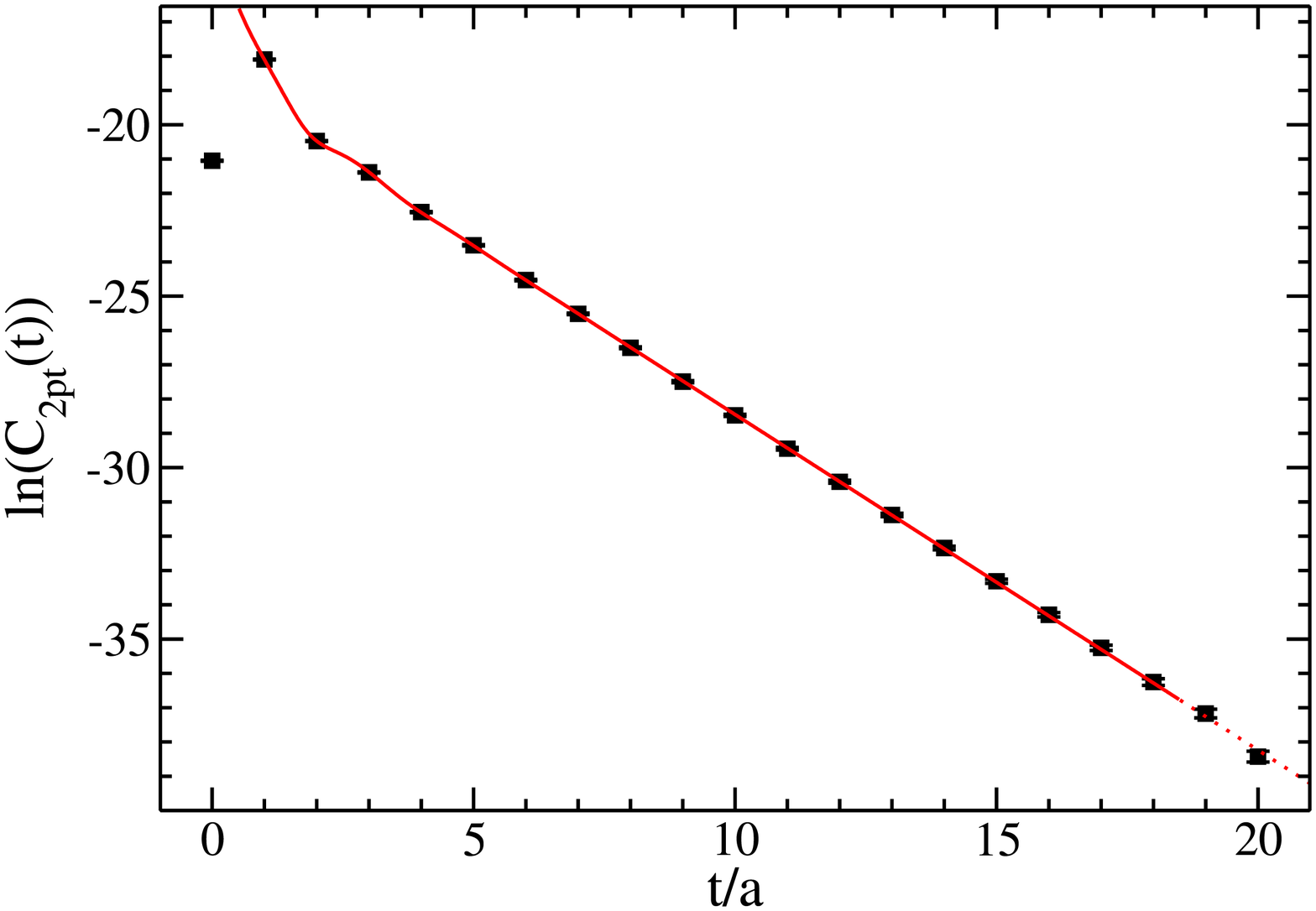}
\caption{\label{2ptfit}Nucleon two-point correlator and fit.}
\end{minipage}
\begin{minipage}{17.75pc}
\includegraphics[width=16.5pc]{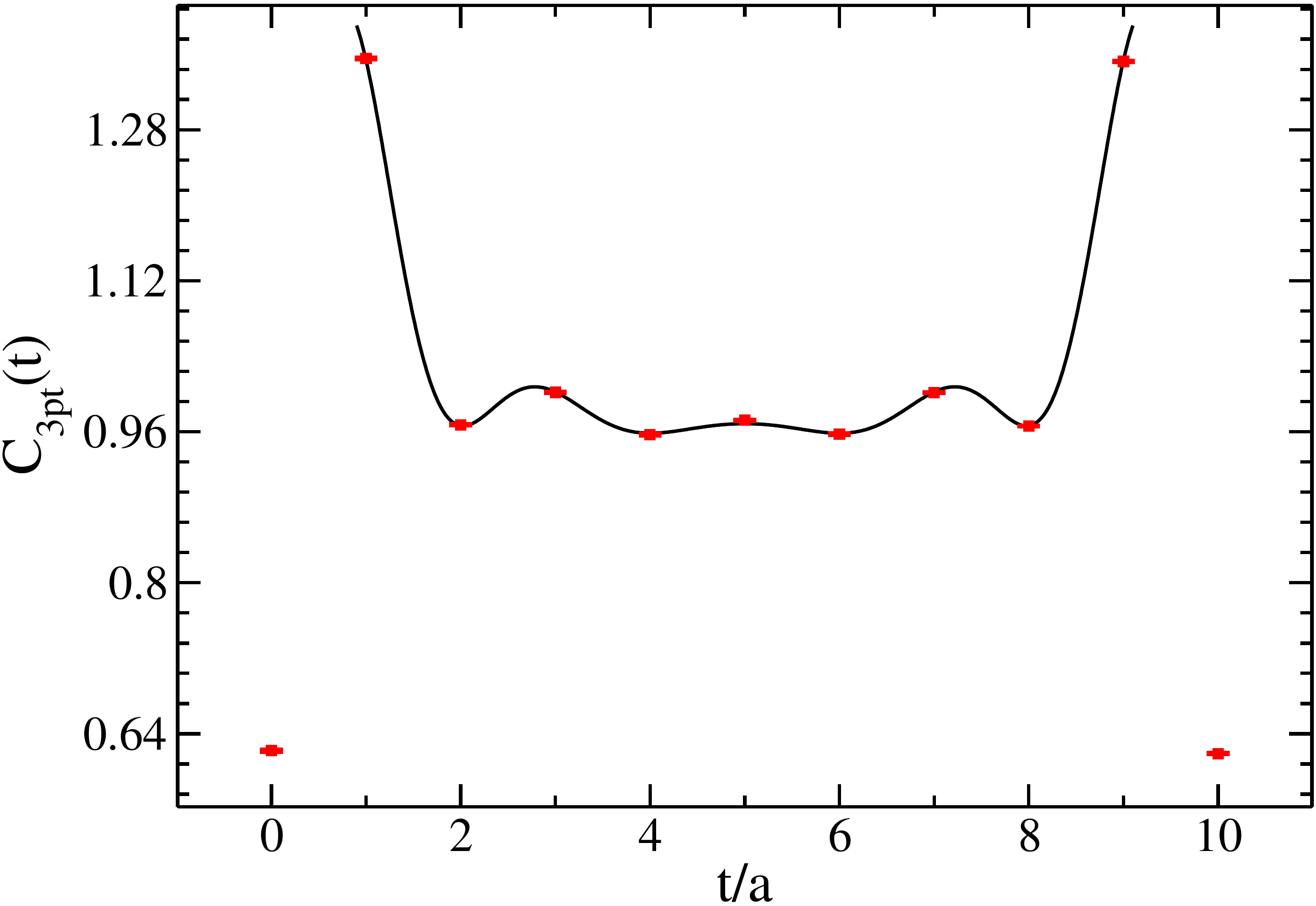}
\caption{\label{3ptfit}Nucleon three-point plateau and fit.}
\end{minipage}
\end{figure}

\begin{figure}
\begin{minipage}{17.75pc}
\includegraphics[width=16.5pc]{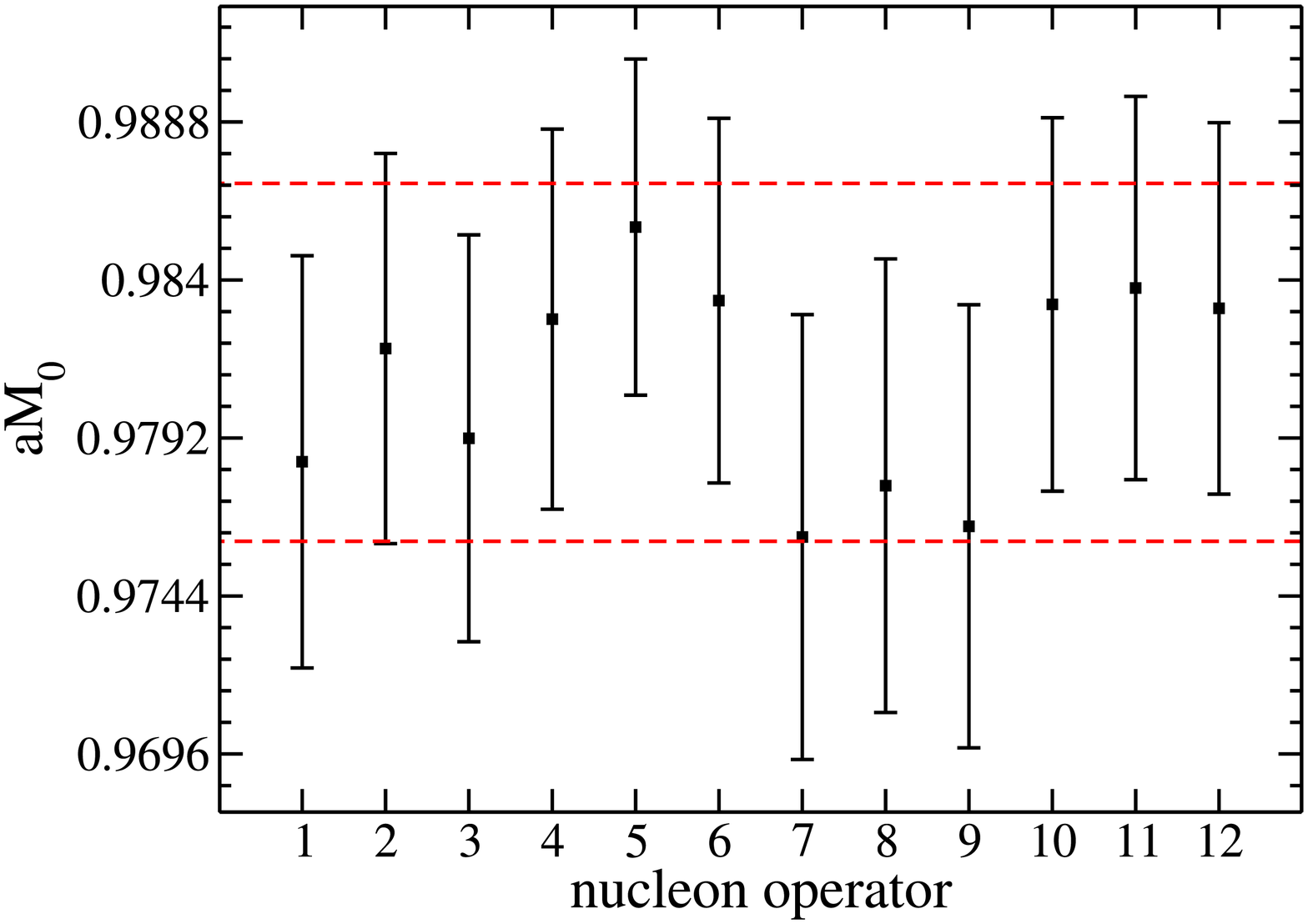}
\caption{\label{M0}Ground state mass for all fits.}
\end{minipage}
\hspace{0pc}
\begin{minipage}{17.75pc}
\hspace{-0.5pc}
\includegraphics[width=16.5pc]{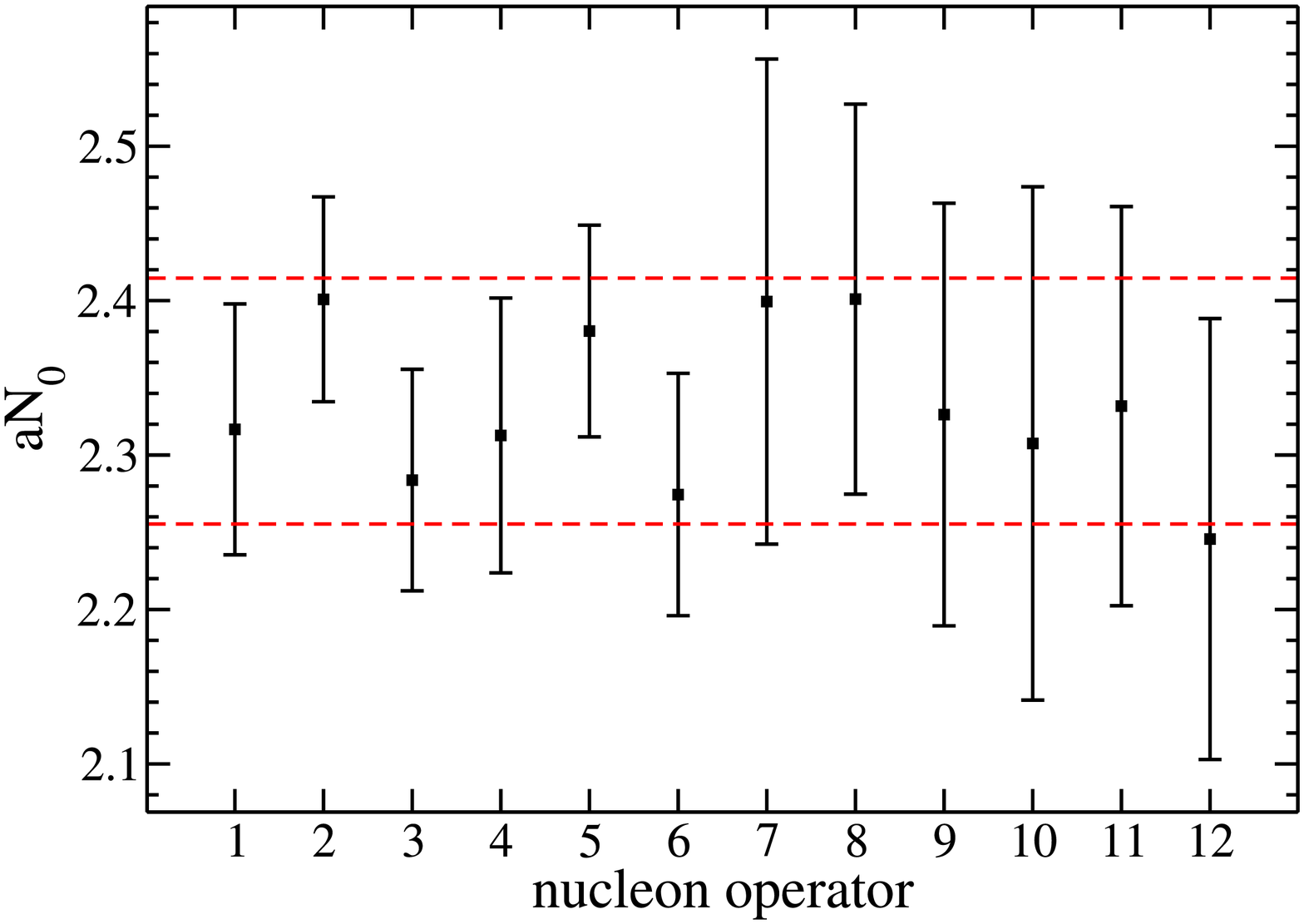}
\caption{\label{N0}Lowest oscillating mass for all fits.}
\end{minipage}
\end{figure}

Simple lattice actions admit a transfer matrix formalism, which allows
one to rigorously relate lattice correlation functions and quantum
mechanical matrix elements.  Domain wall fermions have a transfer
matrix in the full five-dimensional space, however, that does not
guarantee positive definite correlation functions in four
dimensions.~\footnote{Mixed action calculations in general suffer from
a lack of unitarity, despite the choice of valence action.}  This
results in clear oscillations in correlation functions, as shown in
Figs.~[\ref{2ptfit}] and [\ref{3ptfit}].
Lacking a theoretical derivation of an effective four-dimensional
transfer matrix, we consider the following phenomenological forms,
which include the standard positive-definite contributions as well as
oscillating contributions.
\begin{eqnarray}
\label{2ptform}
C_{2\mathrm{pt}}(t) &=& A_0 \exp(-M_0 t) + A_1 \exp(-M_1 t) + B_0 (-1)^t \exp(-N_0 t)\\
\label{3ptform}
C_{3\mathrm{pt}}(t) &=& O_{00} + O_{10} \cosh( (M_1-M_0) t) + P_{00} (-1)^t \cosh( (N_0-M_0) t)
\end{eqnarray}
Figures~[\ref{2ptfit}] and [\ref{3ptfit}] show fits to
Eqs.~[\ref{2ptform}] and [\ref{3ptform}], respectively, for the case
$m_\pi=761~\mathrm{MeV}$.  To test this functional form, we fit 12
nucleon two-point correlation functions.  The resulting ground state
and lowest oscillating state masses are shown in Figs.~[\ref{M0}] and
[\ref{N0}] respectively, again for $m_\pi=761~\mathrm{MeV}$.  There is
clear agreement for $M_0$, $N_0$, and $M_1$ (not shown) for each
nucleon channel considered, indicating universal values for the
physical masses, as expected, and also for the oscillating mass.
Therefore we employ this fitting strategy for all the moments of
parton distributions presented here and will extend this method to the
generalized form factors in a future work.

\subsection{Chiral Perturbation Theory}

There is a variety of methods to perform chiral extrapolations of
moments of parton distributions, which should all agree at light
enough quark masses, but will, however, systematically differ when
applied at the quark masses currently used in lattice calculations.
Lattice calculations at the physical quark masses will of course
eliminate the need for chiral extrapolations, however, in the interim,
we examine a variety of these methods with the ultimate goal of
systematically comparing all such methods.

\begin{figure}
\begin{minipage}{17.75pc}
\includegraphics[width=16.5pc]{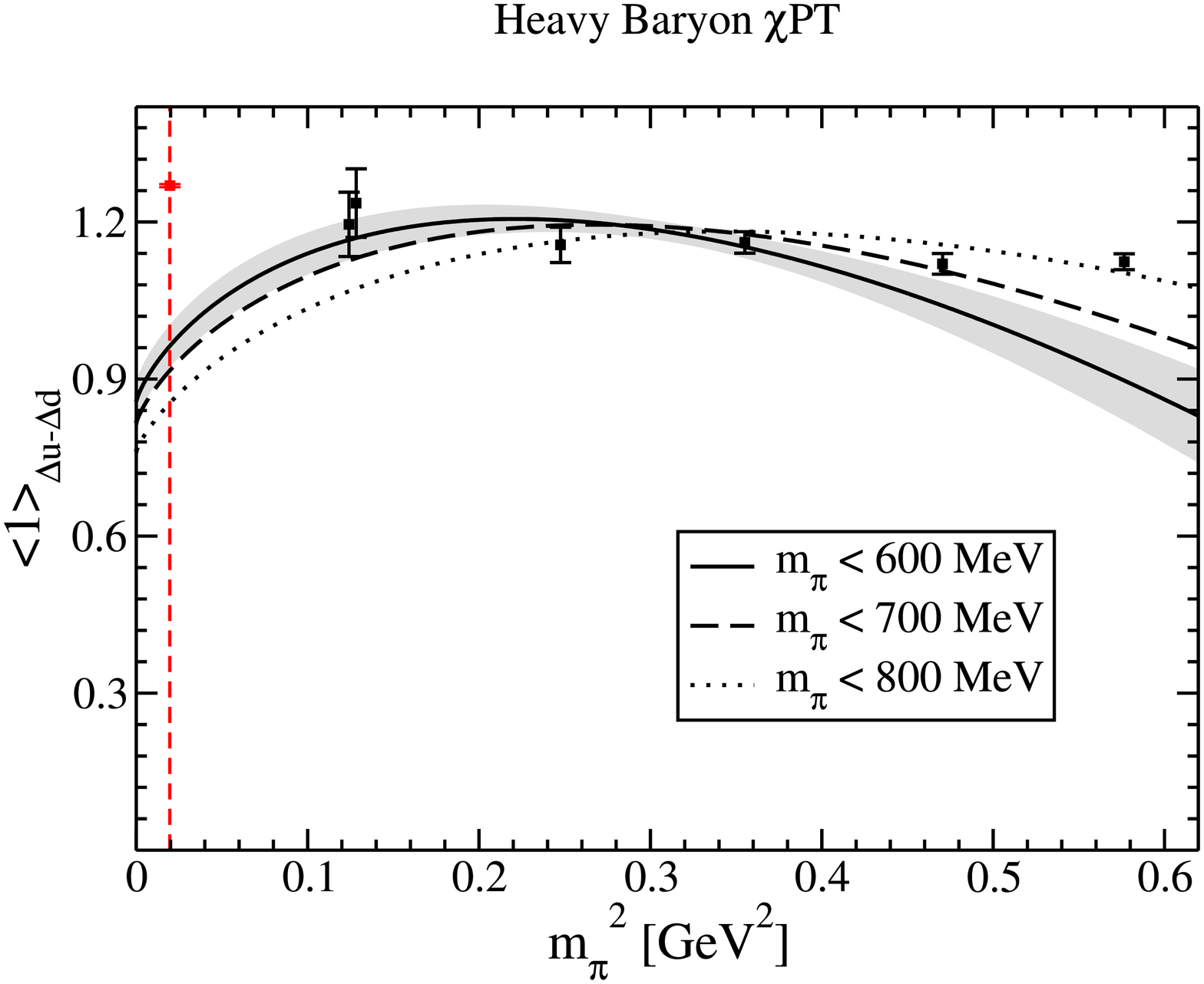}
\caption{\label{gahb}$g_A$ and heavy baryon fit.}
\end{minipage}
\begin{minipage}{17.75pc}
\includegraphics[width=16.5pc]{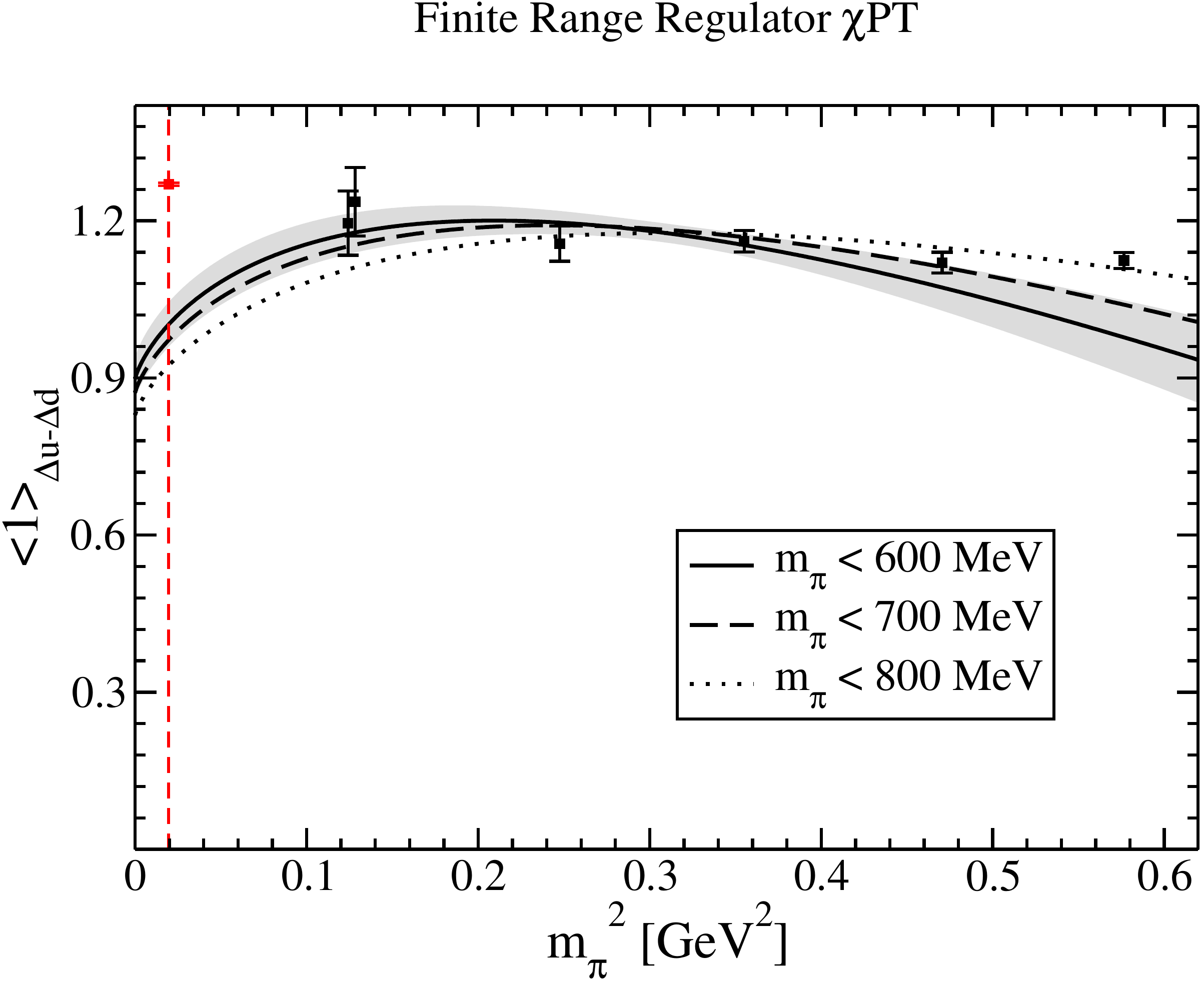}
\caption{\label{gafr}$g_A$ and finite range regulator fit.}
\end{minipage}
\end{figure}

The axial charge and the pion decay constant are low energy constants
that are common to all chiral peturbation theory expressions for the
moments of parton distributions.  Therefore any successful
extrapolation method must account for both observables.  First we
examine two methods that are known to fail.  The first is standard
heavy baryon chiral perturbation theory and the second is the same
expression but with a finite range regulator~\cite{Detmold:2001jb}.
Fits to the axial charge for each are shown in Figs.~[\ref{gahb}] and
[\ref{gafr}].
The figures show the error band for the most conservative fit,
including just the lightest three pion masses.  There is a clear
discrepancy between the experimental result and the extrapolated
lattice result.  Additionally, varying the fit range from
$m_\pi<600~\mathrm{MeV}$ to $m_\pi<800~\mathrm{MeV}$ shows a
systematic variation.

\begin{figure}
\begin{minipage}{17.75pc}
\includegraphics[width=16.5pc]{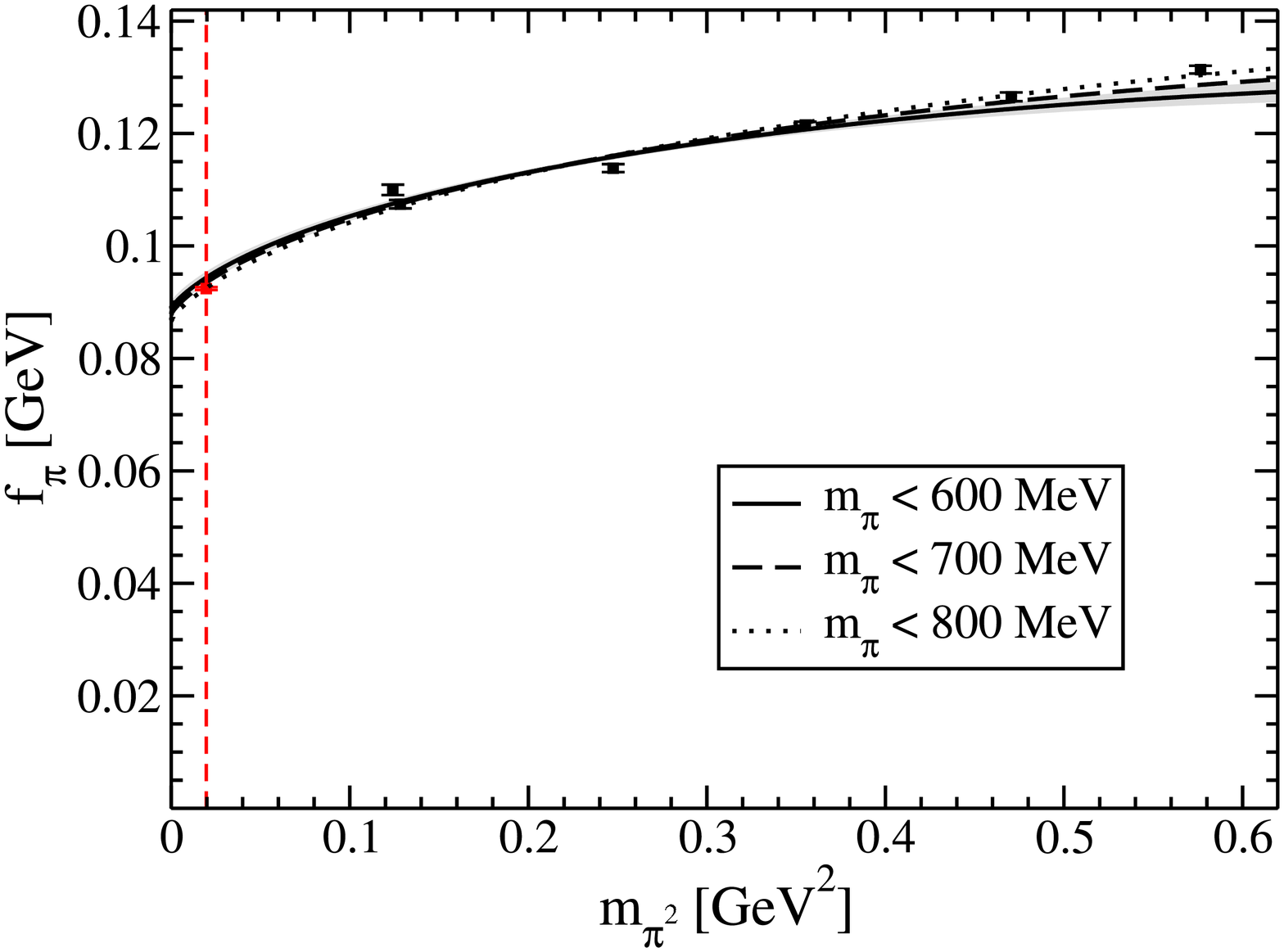}
\caption{\label{fpi}$f_\pi$ and self-consistent fit.}
\end{minipage}
\begin{minipage}{17.75pc}
\includegraphics[width=16.5pc]{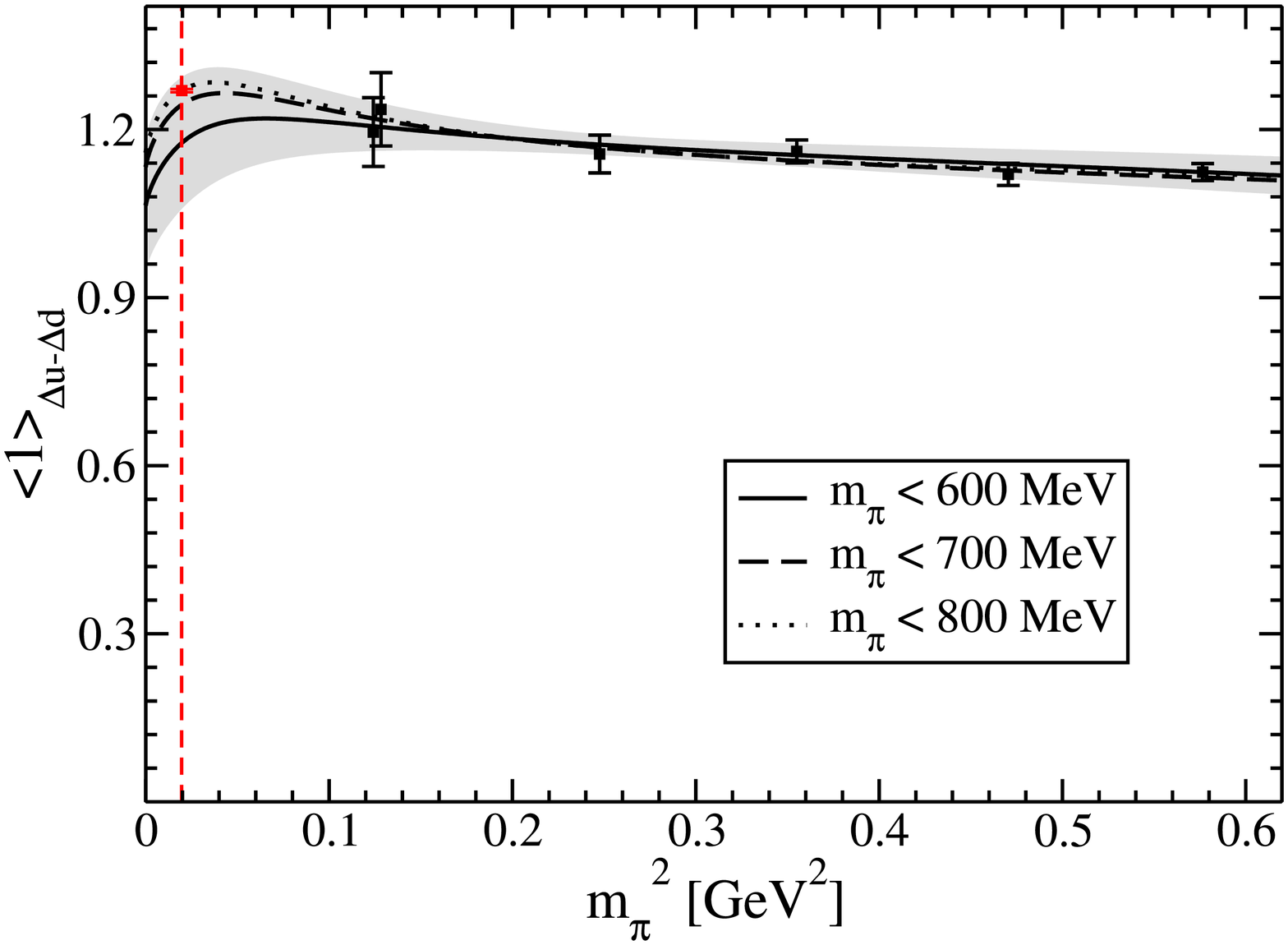}
\caption{\label{ga}$g_A$ and self-consistent fit.}
\end{minipage}
\end{figure}

\begin{figure}
\begin{minipage}{17.75pc}
\includegraphics[width=16.5pc]{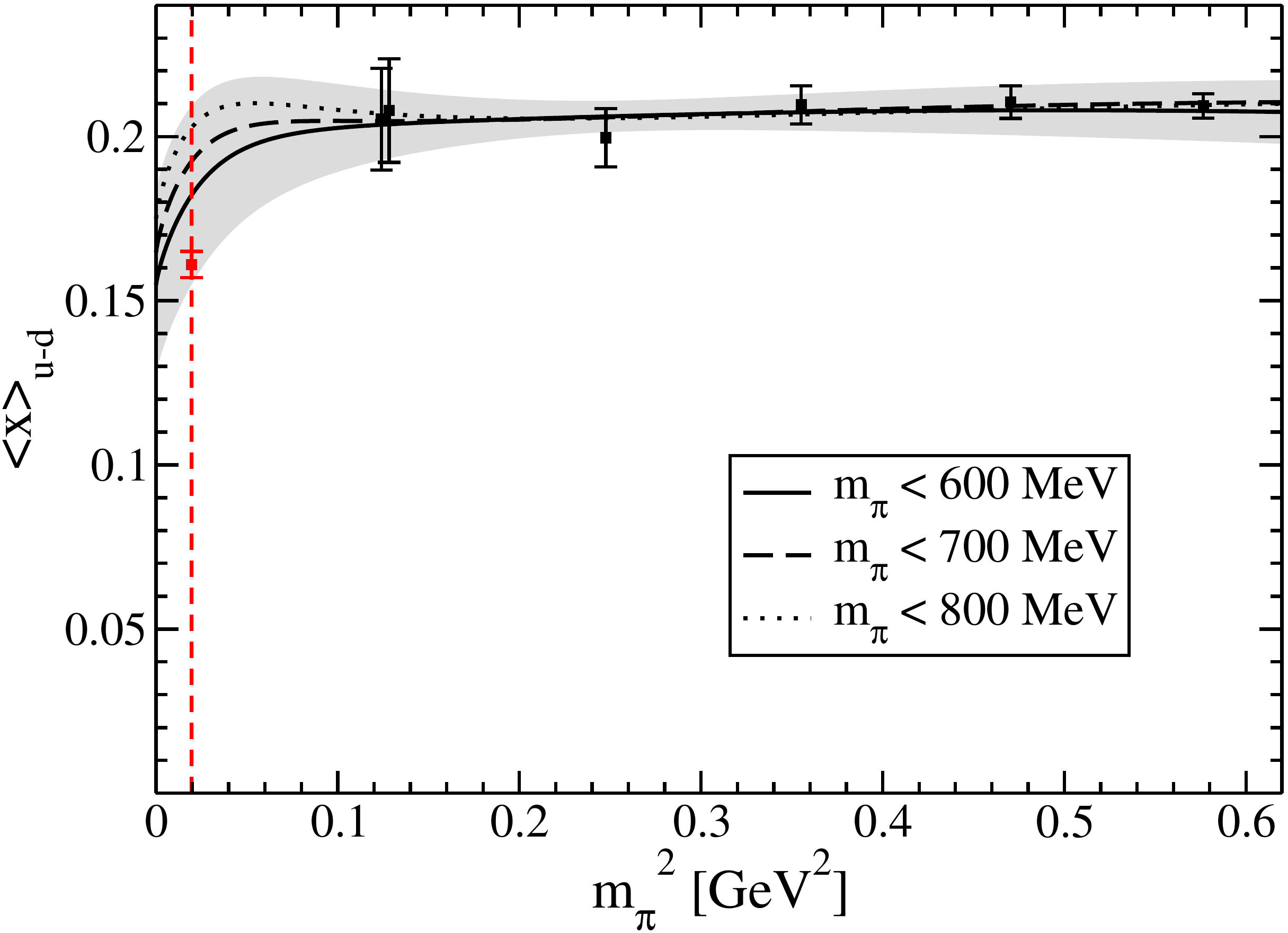}
\caption{\label{xu}$\langle x\rangle_{u-d}$ and self-consistent fit.}
\end{minipage}
\begin{minipage}{17.75pc}
\includegraphics[width=16.5pc]{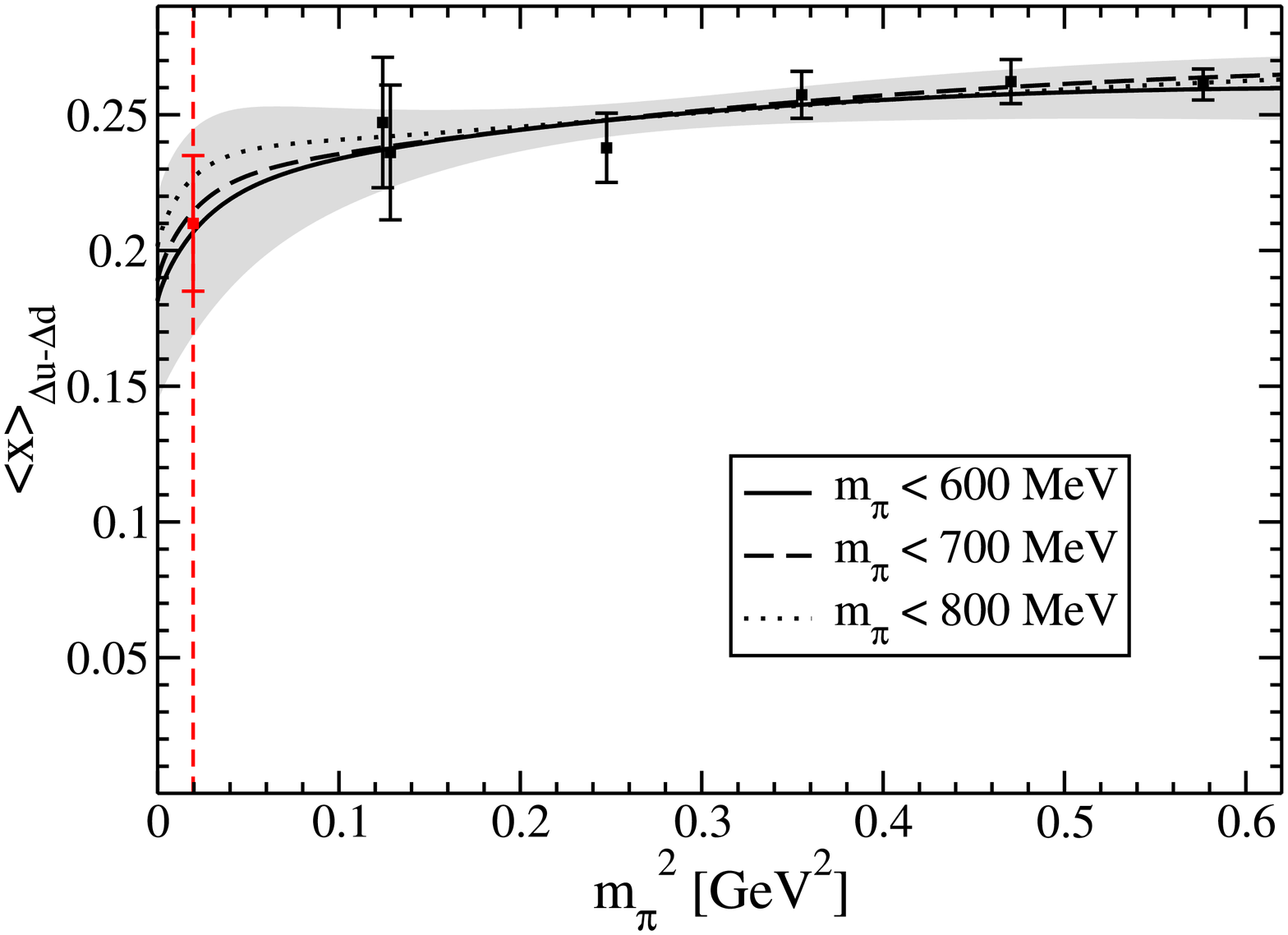}
\caption{\label{xl}$\langle x\rangle_{\Delta u-\Delta d}$ and self-consistent fit.}
\end{minipage}
\end{figure}

The above results indicate that the simplest chiral perturbation
theory expressions fail to converge for pion masses in the range
considered here.  A common alternative is to include the Delta in the
effective theory and this has been shown to successfully describe the
axial coupling in Refs.~\cite{Edwards:2005ym} and \cite{Khan:2006de}.
Here we describe a simpler alternative~\cite{Edwards:2006qx} that
seems to correctly reproduce the experimental values for $g_A$,
$f_\pi$, and other low moments of parton distribution functions.  We
start with the standard chiral perturbation expressions for the
various moments of parton
distributions~\cite{Arndt:2001ye,Chen:2001eg}.  The renormalization
scale $\mu$ is eliminated in favor of a dimensionless quantity,
$\alpha$, by setting $\mu=\alpha \fo$.\footnote{The chiral limit
values of $g_A$, $f_\pi$, and moments like $\langle x\rangle$ are
denoted by $\go$, $\fo$, and $\xo$.  Additionally, explicit labels
of $\pi$ are dropped in Eqs.~[\ref{xsc},\ref{gsc},\ref{fsc}].}  The
values of $g_A$ and $f_\pi$ in the chiral limit occur in the
expressions for the moments, but always in the next-to-leading-order
(NLO) term.  In order to eliminate the need for multiple combined
fits, we can eliminate the chiral limit values of all observables that
occur in the NLO terms in favor of their lattice values at the
corresponding pion mass.~\footnote{For similar ideas, see also
Refs.~\cite{Beane:2005rj,Chen:2005ab,O'Connell:2006sh,Beane:2006gj,Beane:2006kx,Beane:2006fk,Beane:2006pt,Beane:2006mx}.}
This simplifies the fits and does not change the chiral expressions to
the order at which we are working.  The resulting expressions are
given below.
\begin{eqnarray}
\label{xsc}
\langle x \rangle ( 1 + (3 g^2 + 1) m^2/(4\pi f)^2 \ln ( m^2/(\alpha f)^2 ) ) &=&
\xo + c_x(\alpha) m^2\\
\label{gsc}
g ( 1 + (2 g^2 + 1) m^2/(4\pi f)^2 \ln ( m^2/(\alpha f)^2 ) ) &=&
\go + c_g(\alpha) m^2\\
\label{fsc}
f ( 1 + m^2/(4\pi f)^2 \ln ( m^2/(\alpha f)^2 ) ) &=&
\fo + c_f(\alpha) m^2
\end{eqnarray}
Each of Eqs.~[\ref{xsc},\ref{gsc},\ref{fsc}] is an independent simple
linear fit for a chiral limit value and a counter-term.  To plot a
moment as a smooth curve that corresponds to these fits, one first
solves the transcendental Eq.~[\ref{fsc}] for $f$, then the cubic
equation Eq.~[\ref{gsc}] for $g$, and then the linear Eq.~[\ref{xsc}]
for $\langle x\rangle$.  As examples, the resulting fits for $f_\pi$,
$g_A$, $\langle x\rangle_{u-d}$, and $\langle x\rangle_{\Delta
u-\Delta d}$ are shown in Figs.~[\ref{fpi}-\ref{xl}].
In each case we find agreement, within the statistical errors, between
the chiral extrapolation and the experimental measurement.  In
Fig.~[\ref{sum}] we collect these results along with the lowest
transversity moments in order to highlight the genuine potential for
predictions of transversity distributions from lattice QCD
calculations.  In this figure, we normalize each result to the
experimental measurment where available; otherwise we normalize by the
lattice calculation.  For each observable, the leftmost point is the
lattice calculation and the rightmost point is the experimental
measurement.

\section{Generalized Form Factors}

The low non-singlet moments of the nucleon parton distributions serve
as benchmark calculations: as the precision for the moments increases,
so does the confidence in the results for the generalized form
factors.  Extensive results relating to the transverse structure and
spin decomposition as well as several comparative studies of chiral
perturbation are given in Ref.~\cite{Hagler:2007xi}.  Here we focus on
one example of the application of chiral perturbation theory to the
generalized form factor, $C_{20}$.  As for the parton distributions,
there are several varieties of effective field theory methods
available to extrapolate both the pion mass and momentum, $t$,
dependence of generalized form factors.
\begin{figure}
\begin{minipage}{17.75pc}
\includegraphics[width=16.5pc]{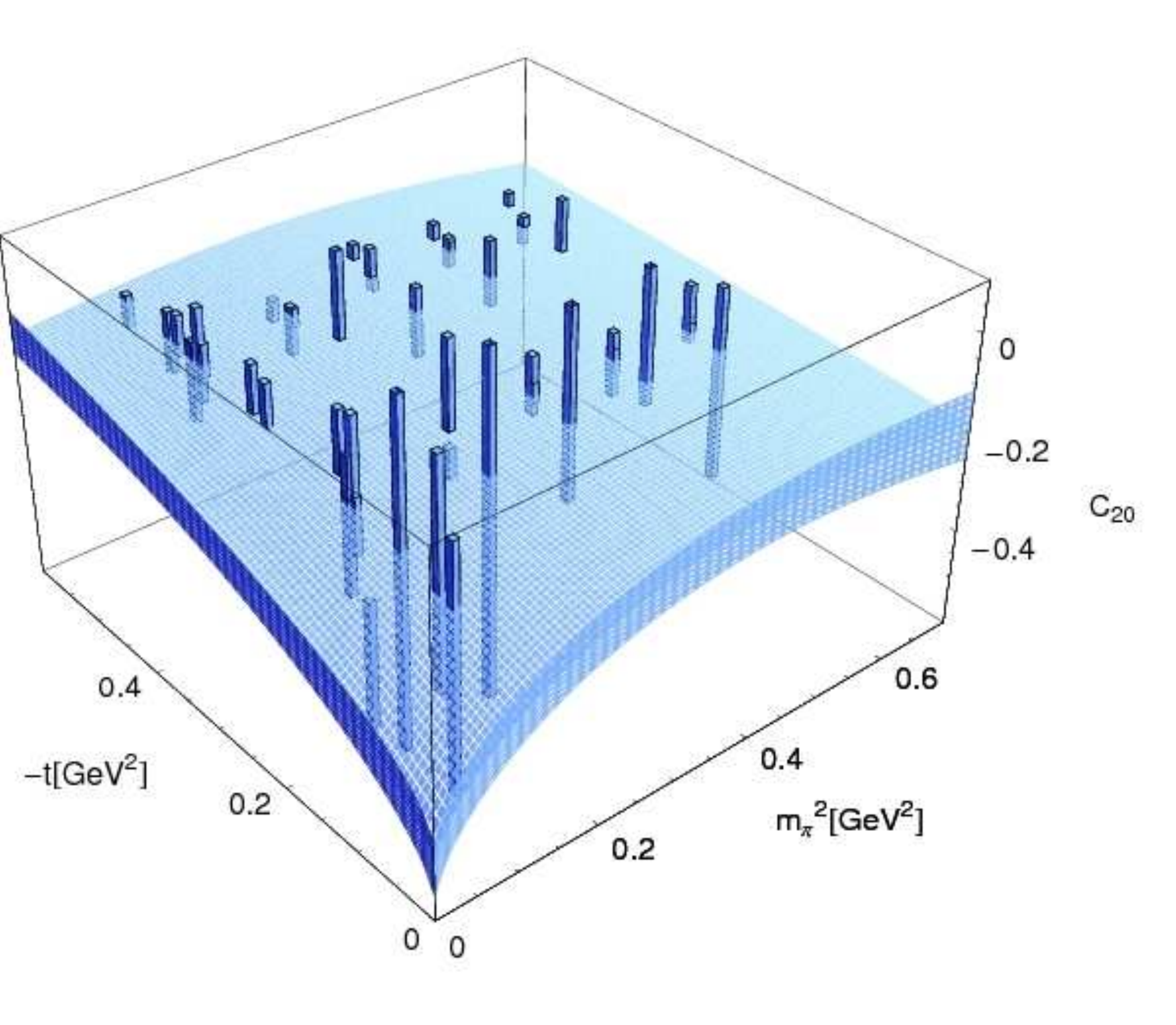}
\caption{\label{C20hb}$C_{20}(t)$ and heavy baryon fit.}
\end{minipage}
\begin{minipage}{17.75pc}
\includegraphics[width=16.5pc]{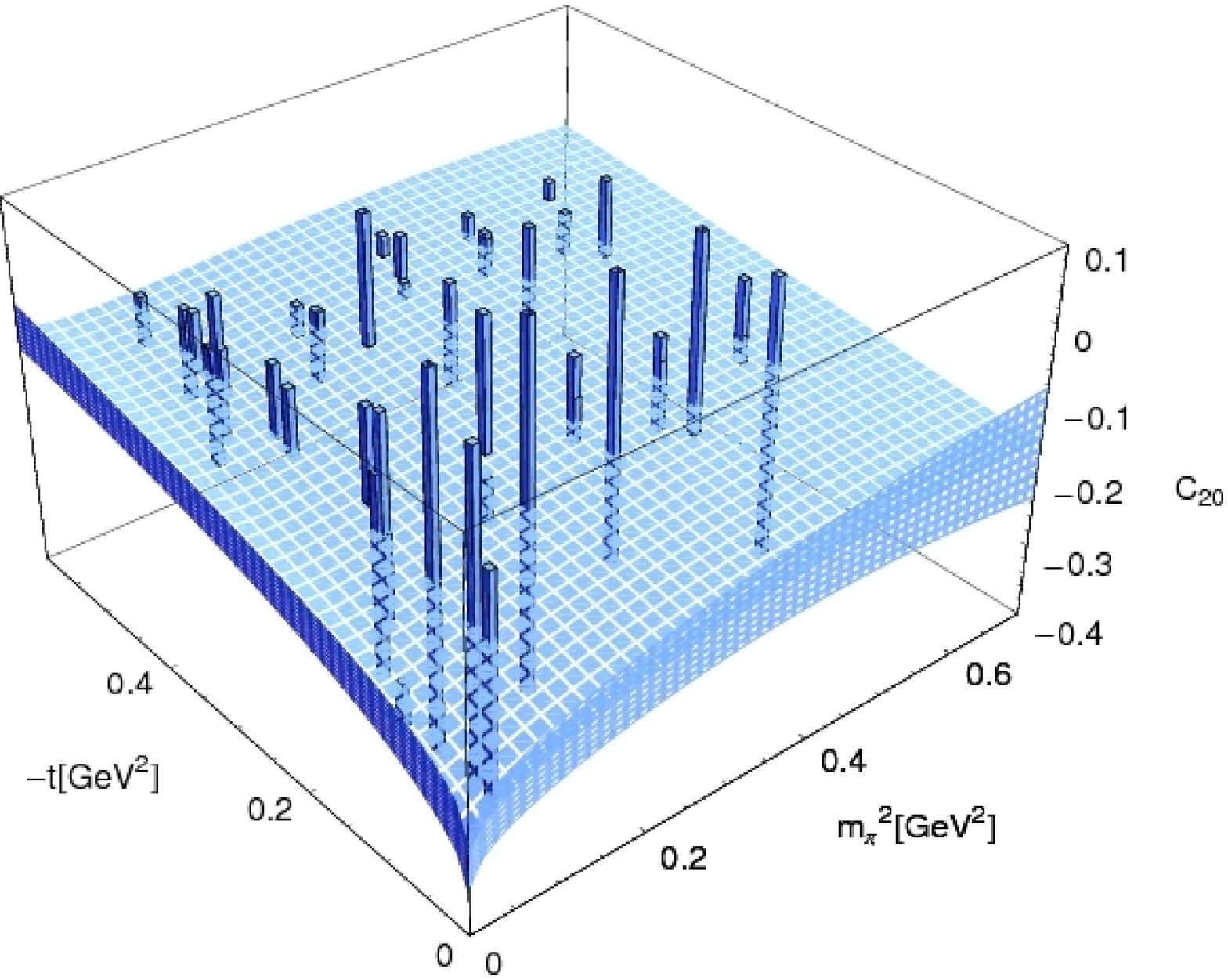}
\caption{\label{C20cov}$C_{20}(t)$ and covariant fit.}
\end{minipage}
\end{figure}
Figures~[\ref{C20hb}] and [\ref{C20cov}] show our results for the
$C_{20}$ form factor and the fits to heavy
baryon~\cite{Diehl:2006js,Diehl:2006ya} and
covariant~\cite{Dorati:2007bk} chiral perturbation theory.  These
results illustrate again the benefits of examining multiple methods of
chiral extrapolations.

\section{Conclusions}

Given the scale of computing resources estimated in the near future,
it is likely that the simplest aspects of nucleon structure will be
calculated with quantitatively controlled errors.  This will require a
concerted effort to control all sources of errors in our calculations.
We examine here the traditional sources of error associated with
extracting matrix elements from lattice correlation functions as well
as several varieties of chiral extrapolation methods.  Though the
simplest methods still fail to correctly accomodate the current
lattice results and reproduce the experimental measurements, we find
that the self-consistent replacement of the chiral limit values of
$g_A$, $f_\pi$, and the moments $\langle x^n\rangle$ with the
correpsonding lattice results appears to improve convergence in all
cases that can be compared with experimental results.  Additionally,
we illustrate the potential to examine not only the pion mass
dependence but also the momentum dependence of form factors using
chiral perturbation theory.

\section{Acknowledgments}

This work was supported by the DOE Office of Nuclear Physics under
contracts DE-FC02-94ER40818, DE-AC05-06OR23177 and DE-AC05-84150, the
EU Integrated Infrastructure Initiative Hadron Physics (I3HP) under
contract RII3-CT-2004-506078, the DFG under contract FOR 465
(Forschergruppe Gitter-Hadronen-Ph\"anomenologie) and the DFG
Emmy-Noether program.  Computations were performed on clusters at
Jefferson Laboratory and at ORNL using time awarded under the SciDAC
initiative. We are indebted to the members of the MILC collaboration
for providing the dynamical quark configurations which made our full
QCD calculations possible.

\bibliography{renner_lat07.bib}

\end{document}